\documentclass[twocolumn,preprintnumbers,amsmath,amssymb,superscriptaddress]{revtex4}
\usepackage{graphicx}
\usepackage{dcolumn}
\usepackage{bm}
\usepackage{soul}
\usepackage{color}
\usepackage{epstopdf}
\usepackage[version=3]{mhchem}
\usepackage{lipsum}
\usepackage[outercaption]{sidecap}
\usepackage{floatrow}
\begin{document}

\title{Applications of mesoporous silica encapsulated gold nanorods loaded doxorubicin in chemo-photothermal therapy}% Force line breaks with \\
\author{Nghiem Thi Ha Lien}
\affiliation{Center for Quantum and Electronics, Institute of Physics, VAST, Dao Tan 10, Hanoi 10000, Vietnam}
\email{halien@iop.vast.vn}
\author{Anh D. Phan}
\affiliation{Faculty of Materials Science and Engineering, Phenikaa Institute for Advanced Study, Phenikaa University, Hanoi 12116, Vietnam}
\email{anh.phanduc@phenikaa-uni.edu.vn}
\affiliation{Faculty of Information Technology, Artificial Intelligence Laboratory, Phenikaa University, Hanoi 12116, Vietnam}

\author{Bui Thi Van Khanh}
\affiliation{College of Science, Vietnam National University (VNU), Hanoi, 334 Nguyen Trai Road, Thanh Xuan, Hanoi, Vietnam}

\author{Nguyen Thi Thuy}
\affiliation{Center for Quantum and Electronics, Institute of Physics, VAST, Dao Tan 10, Hanoi 10000, Vietnam}

\author{Nguyen Trong Nghia}
\affiliation{Center for Quantum and Electronics, Institute of Physics, VAST, Dao Tan 10, Hanoi 10000, Vietnam}

\author{Hoang Thi My Nhung}
\affiliation{College of Science, Vietnam National University (VNU), Hanoi, 334 Nguyen Trai Road, Thanh Xuan, Hanoi, Vietnam}

\author{Tran Hong Nhung}
\affiliation{Center for Quantum and Electronics, Institute of Physics, VAST, Dao Tan 10, Hanoi 10000, Vietnam}

\author{Do Quang Hoa}
\affiliation{Center for Quantum and Electronics, Institute of Physics, VAST, Dao Tan 10, Hanoi 10000, Vietnam}

\author{Vu Duong}
\affiliation{Center for Quantum and Electronics, Institute of Physics, VAST, Dao Tan 10, Hanoi 10000, Vietnam}

\author{Nguyen Minh Hue}
\affiliation{Center for Quantum and Electronics, Institute of Physics, VAST, Dao Tan 10, Hanoi 10000, Vietnam}
\date{\today}% It is always \today, today,

\begin{abstract}
We investigate chemo-photothermal effects of gold nanorods (GNRs) coated using mesoporous silica (\ce{mSiO_2}) loading doxorubicin (DOX). When the mesoporous silica layer is embedded by doxorubicin drugs, a significant change in absorption spectra enable to quantify the drug loading. We carry out photothermal experiments on saline and livers of mice having \ce{GNRs$@$mSiO_2} and \ce{GNRs$@$mSiO_2}-DOX. We also inject the gold nanostructures into many tumor-implanted mice and use laser illumination on some of them. By measuring weight and size of tumors, the distinct efficiency of photothermal therapy and chemotherapy on treatment is determined. We experimentally confirm the accumulation of gold nanostructures in liver.
\end{abstract}

\maketitle

%\tableofcontents

\section{Introduction}
Photothermal effects of gold nanorods have been extensively investigated due to medical applications \cite{1,2,3,4,5,6}. When shined by light, gold nanorods (GNR) absorb radiation and confine electromagnetic energy to  the surface of a gold–dielectric interface. The absorbed energy is efficiently transformed into thermal energy and locally heats to a few hundred degree Celsiuses under illumination of laser light. Since destruction of living cells occrus at temperatures above 43 \ce{^0C}, this light-to-heat conversion process is used for cancer therapy \cite{1,2,3,4,5,6}. In synthesis of gold nanostructures using cetyl trimethylammonium bromide (CTAB), their gold surfaces become unstable and easily aggregate. This aggregation causes the loss of unique optical properties. Moreover, it has been proved that CTAB is a highly toxic cationic surfactant \cite{7,8}. To exploit GNRs for biomedical applications, it is necessary to replace CTAB and coat GNRs by safer and more biocompatible materials/molecules. Recently, extensive efforts have been devoted to encapsulate GNRs with mesoporous silica \cite{9,10}, bio-polymers \cite{11,12}, or other composites \cite{13,14}. These coating materials can be easily combined with drugs or target molecules to functionalize GNRs.

In some biological and photothermal applications, GNRs is preferred rather than gold nanoshells (GNSs) because of two main reasons. First, the size of GNSs having plasmon resonances in the near-infrared region is large ($\sim 150$nm) \cite{28,32}, while the size of GNRs is small ($\leq 80$ nm) \cite{1,2,3,4,5,6}. According to previous works  \cite{33,34}, sub-100 nm nanoparticles are selected for treatments due to their free movement through tissues. Second, it is well known that GNRs have a strong optical absorption cross section and a very small scattering cross section \cite{35,36}. These behaviors are relatively opposite to properties of large size GNSs. Thus, the GNRs are more photothermally efficient than the GNSs \cite{35,36}.

While doxorubicin is an anthracycline antibiotic known as one of the most therapeutic anti-tumor drugs to treat various solid malignant tumors. Doxorubicin intercalates within DNA helix in the cell nucleus. Thereby, this drug inhibits macromolecular biosynthesis and destroys DNA function \cite{21}. However, free DOX has its own several side effects when used in cancer chemotherapy \cite{15}. To reduce the toxicity and side effects and improve treatment efficiency, one encapsulates DOX in liposomes \cite{22}, loads onto silica mesoporous nanoparticles \cite{23}, or incorporates into nanocomposites such as silica mesoporous encapsulated GNRs \cite{24}, GNR@m\ce{SiO_2} or GNR@polymers \cite{9, 12}. Doxorubicin release and loading from GNR@m\ce{SiO_2}-DOX or GNR@polymer-DOX strongly depend on acidic medium and laser irradiation at surface plasmon resonances \cite{9}. The synergistic effect of the drug loaded GNR@\ce{SiO_2} has been demonstrated \cite{9,14,15}.

There have recently been clinical trials for gold-nanoparticle-based photothermal cancer therapy to ablate thermally neck, brain, lung, and prostate cancerous tumors \cite{9,16,17,18}. These results showed promising scenarios since tumors were locally heated. However, accumulation of gold nanostructures in liver and spleen throughout life in bodies without excrete is the most weakest point of this approach \cite{17,18}. The penetration of gold composites into tumors is tiny compared to the infused initial counterparts.

In this work, we synthesize GNRs@m\ce{SiO_2}-DOX and investigate photothermal and chemical behaviors of nanostructures under laser irradiation. We use transmission electron microscopy (TEM) and X-ray diffraction (XRD) spectrum to analyze structures of experimental samples. Effects of DOX loading on optical properties are determined using the absorption spectrum. Then, we inject solutions of GNRs@m\ce{SiO_2} with and without DOX molecules into 45 normal and tumor-planted mice and study chemo-photothermal activities. Based on experimental results, we quantitatively evaluate roles of individual mechanism on cancer treatments.

\section{Experimental Section}
\subsection{Materials}
We purchased tetrachloroauric acid trihydrate 99.5 $\%$ (\ce{HAuCl_4.3H_2O}), ethanol, CTAB, ascorbic acid, and \ce{AgNO_3} from Merck. Tetraethoxylsilane and \ce{NH_4OH} (28-30 $\%$), and \ce{NaBH_4} were purchased from Sigma-Aldrich and Walko, respectively. Doxorubicin hydrochloride 2 mg/ml was provided by Ebewe Pharma. In our experiments, we dispersed materials into double distilled water.

\subsection{Preparation and characterization of \ce{GNR$@$SiO_2}-DOX complex}
The synthesis of \ce{GNR$@$SiO_2}-DOX composites consists of three steps. First, we synthesized near-infrared light-responsive GNRs. Second, a mesoporous silica layer is coated on gold surfaces. A thickness of the silica layer is approximately 15 nm. Third, when dispersing DOX.HCl molecules in an aqueous solution of GNSs, the mesoporous silica surface adsorbed the biomolecules onto the mesoporous silica surface via electrostatic interactions. Finally, we obtained mesoporous silica-encapsulated GNRs (\ce{GNR$@$mSiO_2}-DOX) composites loading DOX molecules.

%\begin{figure}[htp]
%\center
%\includegraphics[width=8.5cm]{Figure1.pdf}
%\caption{\label{fig:1}(Color online) Schematic representation of three main steps in the preparation of \ce{GNR$@$SiO_2} having gold nanorods (core) and mesoporous silica (shell), and doxorubicin molecules adsorbed onto the mesoporous silica shell.}
%\end{figure}
\subsubsection{Preparation of GNRs}
Gold nanorods having dimensions of $\sim$ $39.2 \times 10.7$ nm (an average aspect ratio of 3.7) were synthesized using the seed-mediated growth method \cite{19}. Briefly, the seed was made by stirring 120 $\mu l$ of 25 mM \ce{HAuCl_4} in 10 ml of 100 mM CTAB solution. Then, we quickly added ice-cold \ce{NaBH_4}(60 $\mu l$, 10 mM) to the seed solution and stirred for 2-6 hours. By mixing 100 ml of CTAB 100 mM, 1.5 ml of \ce{HAuCl_4} 25 mM, 450 $\mu l$ of \ce{AgNO_3} 25 mM and 270 $\mu l$ of L-ascorbic acid 100 mM, a growth solution was prepared. After 2 hours of reaction in a mixture of 1000 $\mu l$ of the seed solution and the growth solution, the GNR seeds were synthesized.

\subsubsection{Preparation of mesoporous silica-encapsulated GNRs}
GNRs were prepared by a seed-mediated sequential growth and a reduction of gold salt in the presence of in the presence of a CTAB surfactant. A 20 ml aliquot of the GNRs solution was centrifuged and re-dispersed in 20 ml deionized water. Then, we added 0.5 ml of the TEOS ethanol solution (20 mM) to 20 mL of GNR aqueous solution (pH is adjusted to 10-11 by mixing \ce{NH_4OH}). After vigorously stirring for 24 hours at room temperature, there was a 15nm thick mesoporous silica layer forming on the surface of the GNRs through hydrolysis and condensation of TEOS. The silica-encapsulated GNRs were isolated by centrifugation, washed with deionized water and ethanol several times, and then re-dispersed in 2 ml of deionized water for later use. 

\subsubsection{Mesoporous silica-encapsulated GNRs functionalized with doxorubicin drugs}
We added 200 $\mu l$ of DOX.HCl 2-mg/mL solution to a solution of \ce{GNRs$@$mSiO_2} solution which has the optical density (OD) equal to 12. Then, this mixture was stirred and mixed during 75h at room temperature in the dark. Electrostatic interaction between doxorubicin molecules and the mesoporous silica surface leads to molecular adsorption at particle surfaces \cite{25,26}. Finally, we used UV-VIS-NIR spectrometer to monitor the adsorbing procedure of DOX into mesoporous silica surface.

\subsubsection{Measurements}
In the final GNR suspension, \ce{GNRs$@$mSiO_2} and \ce{GNRs$@$SiO_2}–DOX were characterized by an ultraviolet/visible wavelength (UV-VIS-NIR) spectrophotometer (Shimadzu 2600) and dynamic light scattering. We examined sizes and shapes of GNRs by high resolution transmission electron microscopy (HRTEM, JEM2100-JEOL). Samples were prepared by placing one drop of the GNR suspension on a 200-mesh, copper grid with carbon (SPI Supplies, West Chester, Pennsylvania), and drying in a vacuum oven overnight. We used X-Ray diffraction to determine the chemical composition and crystalline of the obtained GNRs. %We used a Hitachi S-4800-II scanning electron microscope to snapshot \ce{GNRs$@$SiO_2}.

\subsection{In-vivo experiments}
For in vivo experiments, we used healthy male Swiss Albino mice (8–10 weeks old) with weight of 18$\pm$ 1.5 g. The mice had free access to food and water and were maintained on a 14/10 h light/dark cycle. The sarcoma cancer cells were subcutaneously implanted (2$\times10^6$ cells/100 $\mu l$ of Hank’s solution) into the right side of mouse back. The tumor size was daily measured. These animals were examined when their average tumor size grew up to 5-6 mm diameter. We randomly divided forty five mice into nine groups (five mice per groups): one control group and eight experimental groups with two different concentrations of \ce{GNR$@$mSiO_2} and \ce{GNR$@$mSiO_2}-DOX having optical density of 5.5 and 9.2 at 808 nm under 808-nm laser irradiation. Particularly, the groups are

- Control group: without tumor and no treatment.

- Group 1: saline injection.

- Group 2: saline injection + 808-nm laser irradiation; 

- Group 3: \ce{GNR$@$mSiO_2} OD = 9.2.

- Group 4: \ce{GNR$@$mSiO_2}-DOX OD = 9.2

- Group 5: \ce{GNR$@$mSiO_2} OD = 5.5 + 808-nm laser irradiation.

- Group 6: \ce{GNR$@$mSiO_2} OD = 9.2 + 808-nm laser irradiation.

- Group 7: \ce{GNR$@$mSiO_2}-DOX OD = 5.5 + 808-nm laser irradiation.

- Group 8: \ce{GNR$@$mSiO_2}-DOX OD = 9.2 + 808-nm laser irradiation. 

We used a 1 ml syrring 26G needle to inject directly 30 $\mu l$ of phosphate-buffered saline (PBS) 1x or \ce{GNR$@$mSiO_2}/\ce{GNR$@$SiO_2}-DOX solutions into each tumor center of mice. The animals were irradiated with laser (808 nm, 3.25 \ce{W/cm^2}, 1 minute, a spot size of 6 mm) immediately after tumor-direct injection. The laser was focused on the center of the tumor. We measured the temperature rise of tumor surface by an infrared (IR) camera (Chauvin Arnoux C.A 1995 IR camera IP 54). After treatment, we observed health and weight of mice. After 1 month, all mice were euthanized. We dissected their tumors and measured their weight. Average body weights of nine groups of mice as a function of time were shown in Figure S1 in Supporting Information. At that time, we also analyzed blood chemistry and hematology. All experiments were performed in compliance with the policy on animal use and ethics.
\section{Results and discussions}
\begin{figure}[htp]
\center
\includegraphics[width=8.2cm]{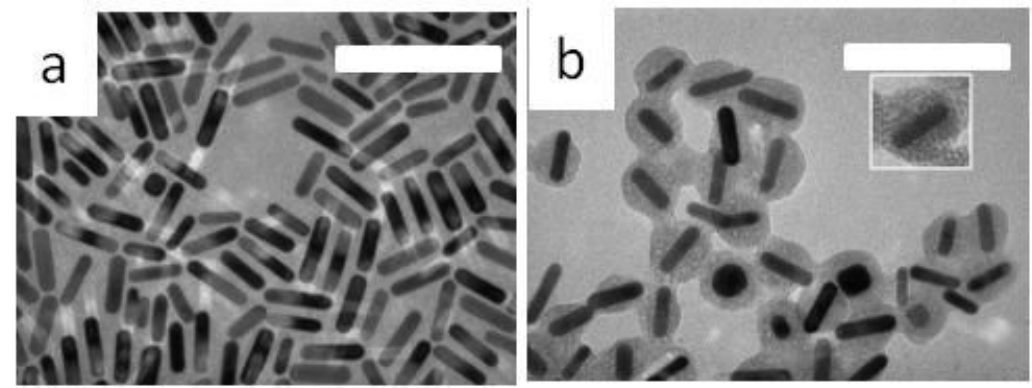}
\includegraphics[width=9cm]{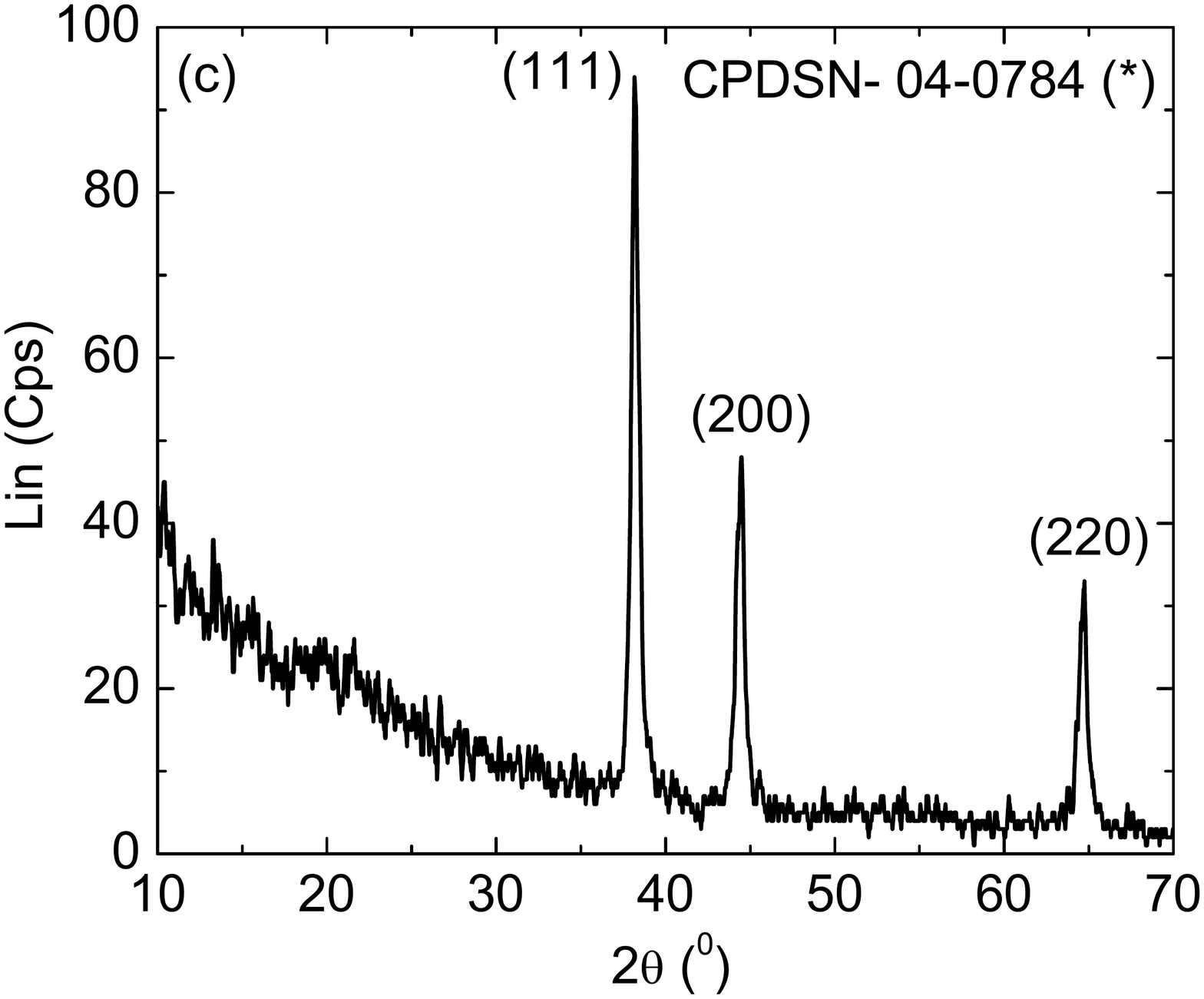}
\caption{\label{fig:2}(Color online) TEM image of (a) GNRs and (b) GNRs@mSiO2 with scale bar of 100 nm, and (c) X-ray diffraction spectrum of GNRs.}
\end{figure}

Figure \ref{fig:2}a and \ref{fig:2}b show TEM images of a solution of GNRs and \ce{GNR$@$mSiO_2}, respectively. The TEM images indicate that the average length and width are 39.2$\pm$ 6.2 nm and 10.7$\pm$1.7  nm, respectively, (the average aspect ratio of 3.7). The silica shell is uniform and has thickness of $\sim$ 15 nm. One can see a TEM image of the mesoporous silica coating on GNRs in Figure S2 in Supporting Information.

Figure \ref{fig:2}c shows X-ray diffraction of GNRs to determine their chemical composition and crystalline. We find three characteristic diffraction peaks at $38.27^0$, $44.42^0$, and $64.70^0$ corresponding to the (111), (200), and (220) crystal planes of the face-centered cubic structure of gold (JCPDS no. 04-0784), respectively. The presence of these optical maxima indicates a polycrystalline structure of GNRs. In addition, since there is no other impurities in the diffraction patterns, our samples are composed of pure crystalline gold.

\begin{figure}[htp]
\center
\includegraphics[width=9cm]{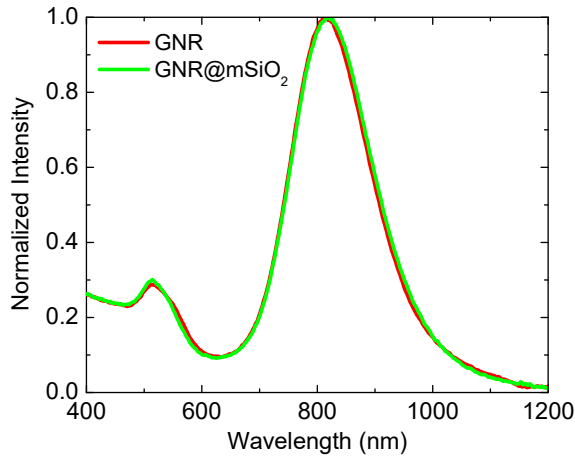}
\caption{\label{fig:4}(Color online) Normalized absorption spectra of GNRs and \ce{GNR$@$mSiO_2}.}
\end{figure}

Figure \ref{fig:4} shows normalized absorption spectra of GNRs (or CTAB-GNRs) and \ce{GNR$@$mSiO_2}. Note that GNRs have CTAB molecules on the surface to stabilize and avoid aggregation. In each spectrum, two peaks corresponds to the transverse ($\sim$513 nm) and longitudinal ($\sim$820 nm) oscillation. The surface plasmon resonance of GNRs is strongly dependent on their aspect ratio \cite{36}. An increase of the aspect ratio leads to a red-shift of the optical peak. However, our results indicates the silica layer has a minor effect on the absorption spectrum. This may be due to a broad distribution of the rod length and width of GNRs as shown in Figure S3 in Supporting Information. This behavior is completely consistent with experimental results in Ref. \cite{31}.

Compared to the optical spectrum of CTAB-GNRs, the longitudinal surface plasmon resonance band of \ce{GNR$@$mSiO_2} is slightly red-shifted about 6 nm. Meanwhile, the peak location for the transverse plasmon band of CTAB-GNRs is at 513 nm and this resonance is blue-shifted about 4 nm when coating silica on the gold surface. This behavior can be explained that the refractive index of silica shell (1.45) is closer to that of water media (1.33) than that of 2-3 nm CTAB layer (1.49). It means the silica layer provides more transmission and lower reflection of electromagnetic fields than the CTAB layer. Thus, more electrons of the GNR core are excited to collectively resonate. The shift of optical resonances as increasing the refractive index of the medium surrounding the core is predictable.

\begin{figure}[htp]
\center
\includegraphics[width=9cm]{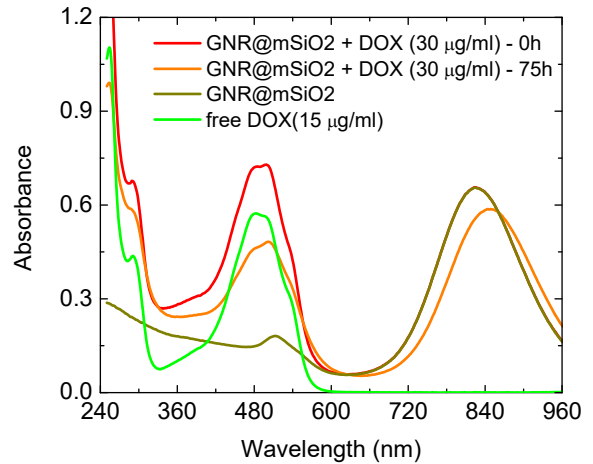}
\caption{\label{fig:5}(Color online) Absorption spectra of free DOX at concentration 30 $\mu g$/ml, \ce{GNR$@$mSiO_2}, \ce{GNR$@$mSiO_2-DOX} (30 $\mu g$/ml) at 0 hour and after 75 hours of incubation.}
\end{figure}

The DOX loading of \ce{GNRs$@$mSiO_2} can be analyzed using the absorption spectra. As shown in Figure \ref{fig:5}, the absorption spectra of \ce{GNR$@$mSiO_2} is blue-shifted after adsorbing DOX. Doxorubicin molecules in the mesoporous silica layer enhance the absorbance intensity of the transverse peak near 500 nm but do not affect the absorbance of the longitudinal peak located at about 813 nm. This is because DOX molecules strongly absorb light at 500 nm and do not interact with NIR radiation. After 75 hours incubation, the optical spectrum of \ce{GNR$@$mSiO_2}-DOX is lowered in comparison with the spectrum measured immediately before and after mixing DOX. These results indicate that DOX molecules tightly bind to the silica layer of \ce{GNR$@$mSiO_2} but biological degradation may appear \cite{29}.

Free and absorbed DOX molecules can be separated using centrifugation. We removed the supernatant with a pipet and determined the concentration of the residual DOX in the supernatant by measuring its absorbance and then comparing this measured spectrum with that of known concentrations of free DOX solution. From this, we estimated that 1 $\mu g$ of Au approximately absorbs 1.3 $\mu g$ of DOX.

\begin{figure}[htp]
\center
\includegraphics[width=9cm]{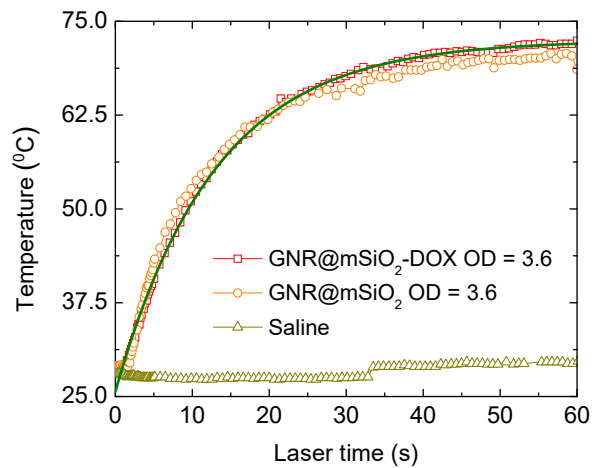}
\caption{\label{fig:6}(Color online) Time-dependent temperature of \ce{GNR$@$mSiO_2}, \ce{GNR$@$mSiO_2}-DOX solutions with optical density of 3.6 and saline irradiated with the NIR laser (808 nm, 3.25\ce{W/cm^2)}. The solid curve is an analytic function $T(t)=T_0+\frac{A}{B}\left(1-e^{-Bt}\right)$ with $T_0=24.67$ $^0C$, $A \approx 3.168$ $^0Cs$, and $B=0.0773$ $s^{-1}$.}
\end{figure}

Under NIR 808-nm laser irradiation with power density at 3.25 \ce{W/cm^2} for 1 minute, \ce{GNRs$@$mSiO_2} and \ce{GNRs$@$mSiO_2}-DOX in solutions absorb the light energy and convert it into heat. Solutions of plasmonic nanocomposites are fixed at the same optical density of 3.6 at 808 nm. During the laser illumination, as shown in Figure \ref{fig:6}, temperature of \ce{GNR$@$mSiO_2} and \ce{GNR$@$mSiO_2}–DOX solutions rapidly increases from room temperature to approximately 70.5 $^0$C. Theoretically, by using the energy balance equation in homogeneous and open systems, one derives an analytical fitting function $T(t)=T_0+\frac{A}{B}\left(1-e^{-Bt}\right)$ to describe the time-dependent temperature \cite{28,30}. The analysis agrees quantitatively well with our experimental data. The temperature rises of these solutions relatively overlap. This result can be deduced from Figure \ref{fig:5} since the presence of DOX molecules does not alter the absorbance of \ce{GNR$@$mSiO_2} in the NIR region. Thus, aqueous solutions of \ce{GNR$@$mSiO_2} and \ce{GNR$@$mSiO_2}–DOX absorb energy and are heated in the same manner. While the temperature is nearly unchanged in saline solution.

Furthermore, since the absorption spectrum of GNRs with and without a mesoporous silica layer on the gold surface have a very small variation (see Figure \ref{fig:4}), one could expect a perfect overlapping between the time-dependent temperature rise of GNR solution and that of \ce{GNR$@$mSiO_2} solution.

\begin{figure}[htp]
\center
\includegraphics[width=8.2cm]{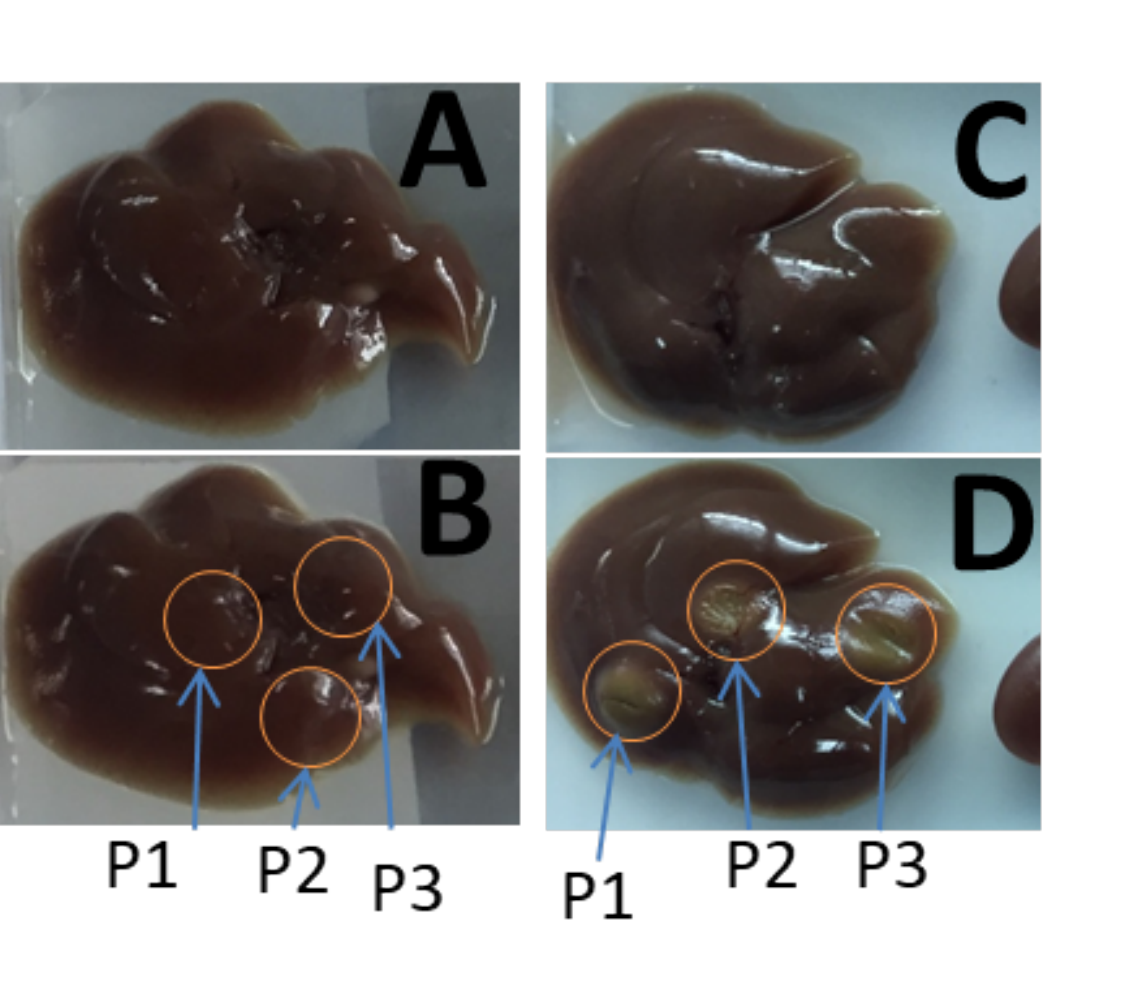}
\includegraphics[width=9cm]{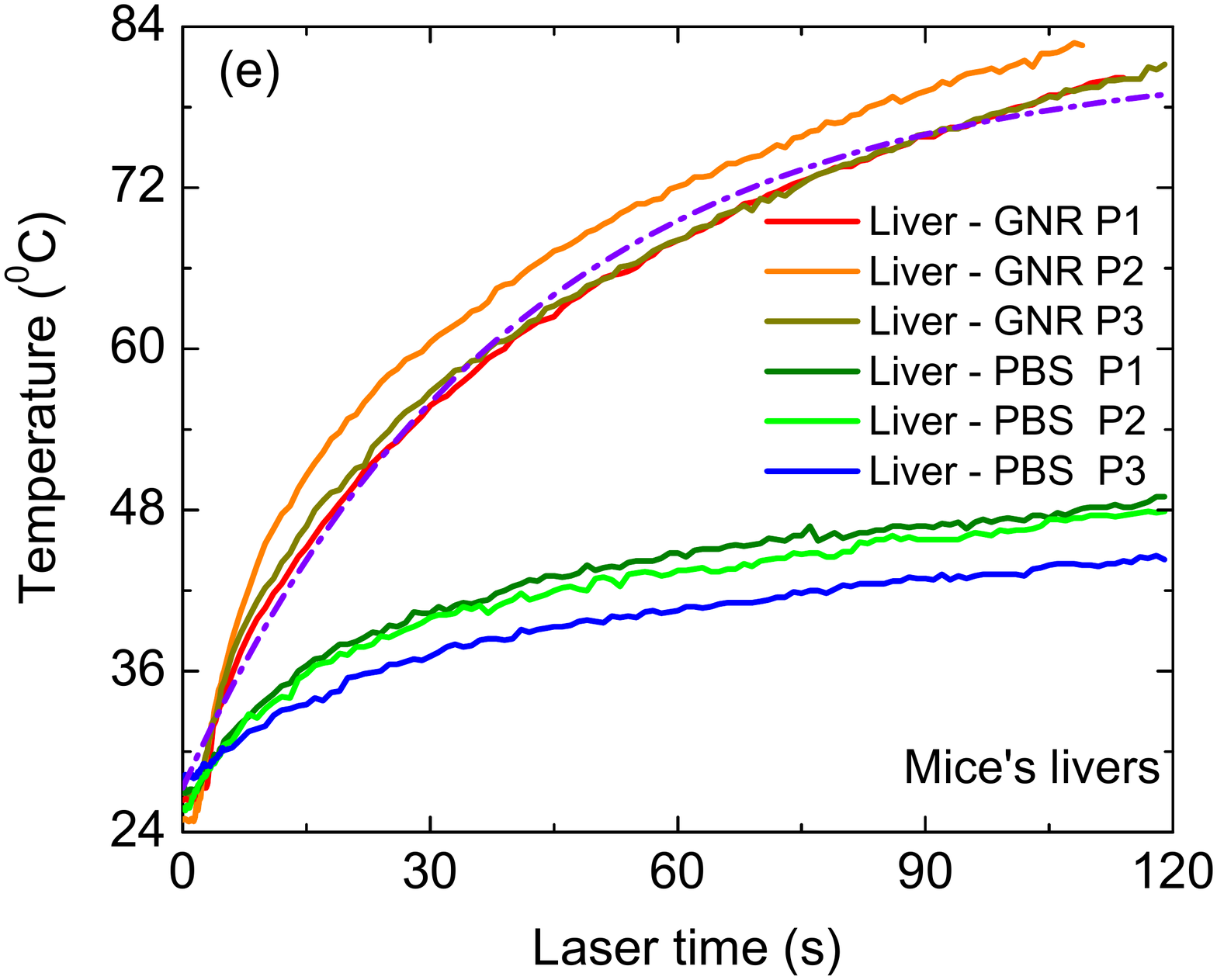}
\caption{\label{fig:7}(Color online) The liver of control mouse image without GNR infusion image before (a) and after (b) irradiated by NIR laser irradiation (808 nm, 3.5 \ce{W/cm^2}, 2 minutes). The liver of mouse image with 150 $\mu l$, OD = 10 of \ce{GNR$@$mSiO_2} infusion before (c) and after (c) NIR laser radiation (808 nm, 3.25 \ce{W/cm^2}, 2 minutes). (e) Temperature rises of mice liver at the three different positions after vein tail injection 2 hours in (b) and (d). The dashed-dotted violet curve is the analytic function $T(t)=T_0+\frac{A}{B}\left(1-e^{-Bt}\right)$ with $T_0=27.29$ $^0C$, $A \approx 1.36$ $^0Cs$, and $B=0.025$ $s^{-1}$.}
\end{figure}

%When low doses of \ce{GNR$@$mSiO_2} (GNR amount of 1.75 \ce{mg. kg^{-1}}) are injected into mice via intravenous injection, . 

Examing chemo-photothermal effects of nanocomposites on animals is an essential part to step towards practical applications. In addition to investigations of the time-dependent temperature rise, the liver accumulation of nanostructures has been intensively concerned in medical-related fields. Several recent works used the dose for intravenous administration of GNRs ranging from 1.7 \ce{mg.kg^{-1}} to 25 \ce{mg.kg^{-1}} \cite{9,27}. It was proved that the amount of GNRs is accumulated in internal organs, particularly in liver and  spleen after infusion. In the remaining life, GNRs cannot be excreted from their bodies \cite{17}. Another study showed that gold nanoparticles are mainly trapped in the Kuffer macrophage cells in the liver for 2 hours after intravenous injection \cite{20}.

In our work, to test the accumulation of \ce{GNR$@$mSiO_2} in the liver, we used low doses (100 $\mu l$ of saline buffer and \ce{GNR$@$mSiO_2} 1.75 \ce{mg.kg^{-1}}) to inject into mice. For two hours after intravenous injection, mice were sacrificed and their livers were taken. The presence and influences of GNRs in the livers were determined via photothermal effects. Under the same conditions of \emph{in vivo} experiments, we illuminated livers at the three different positions ($P_1$, $P_2$, and $P_3$ as depicted in  Figure \ref{fig:7}b and \ref{fig:7}d) within 2 min of NIR laser (808 nm, 3.25 \ce{W/cm^2}). For livers without GNRs, Figure \ref{fig:7}a and \ref{fig:7}b show no difference (visual) between before and after laser radiation at three irradiated positions. Meanwhile, for livers of mice with tail vein injection of \ce{GNR$@$mSiO_2} solutions, one visually observes color change of the liver after laser radiation in Figure \ref{fig:7}d. Color at these three positions irradiated by laser turns yellow compared to Figure \ref{fig:7}c. This indicates the presence of gold nanostructures in the liver and the accumulation leads to the heat-induced denaturation of organs. The time-dependent temperature rises of livers at positions $P_1$, $P_2$, and $P_3$ in Figure \ref{fig:7}b and d, which correspond samples injected by saline buffer and \ce{GNR$@$mSiO_2}, respectively, were monitored and shown in Figure \ref{fig:7}e. The temperatures of experimental samples with \ce{GNRs$@$mSiO_2} increase up to 85$\pm$2 $^0C$ after 2 minutes. For liver with only saline, the temperature slowly increases and reaches to 44$\pm$1 $^0C$. There is no much difference between temperatures at the irradiated points. This finding suggests that the liver is an approximately homogeneous medium and \ce{GNRs$@$mSiO_2} are randomly distributed. Again, the mathematical form of the time-dependent temperature obeys the fit equation $T(t)=T_0+\frac{A}{B}\left(1-e^{-Bt}\right)$.

\begin{figure}[htp]
\center
\includegraphics[width=8cm]{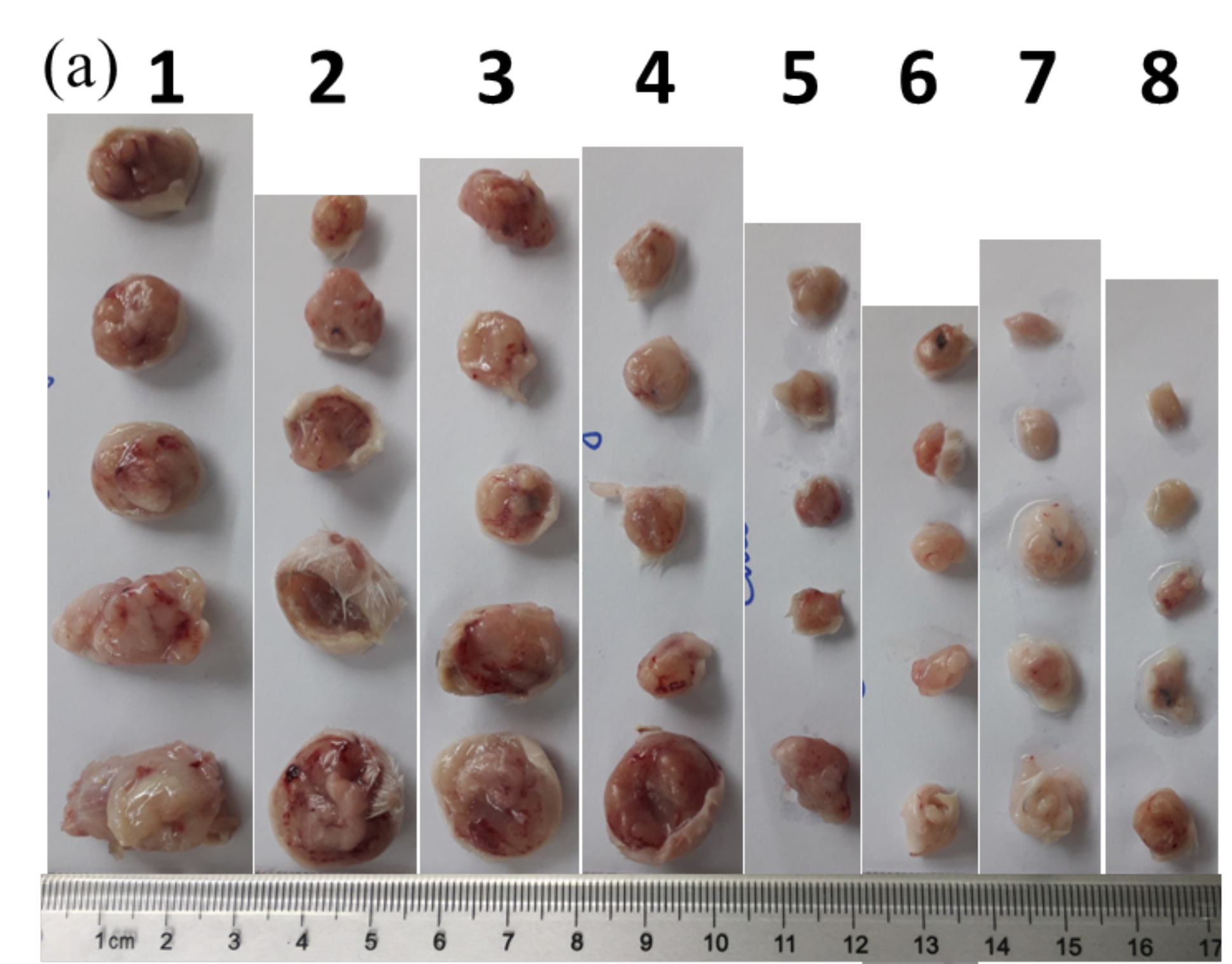}
\includegraphics[width=9cm]{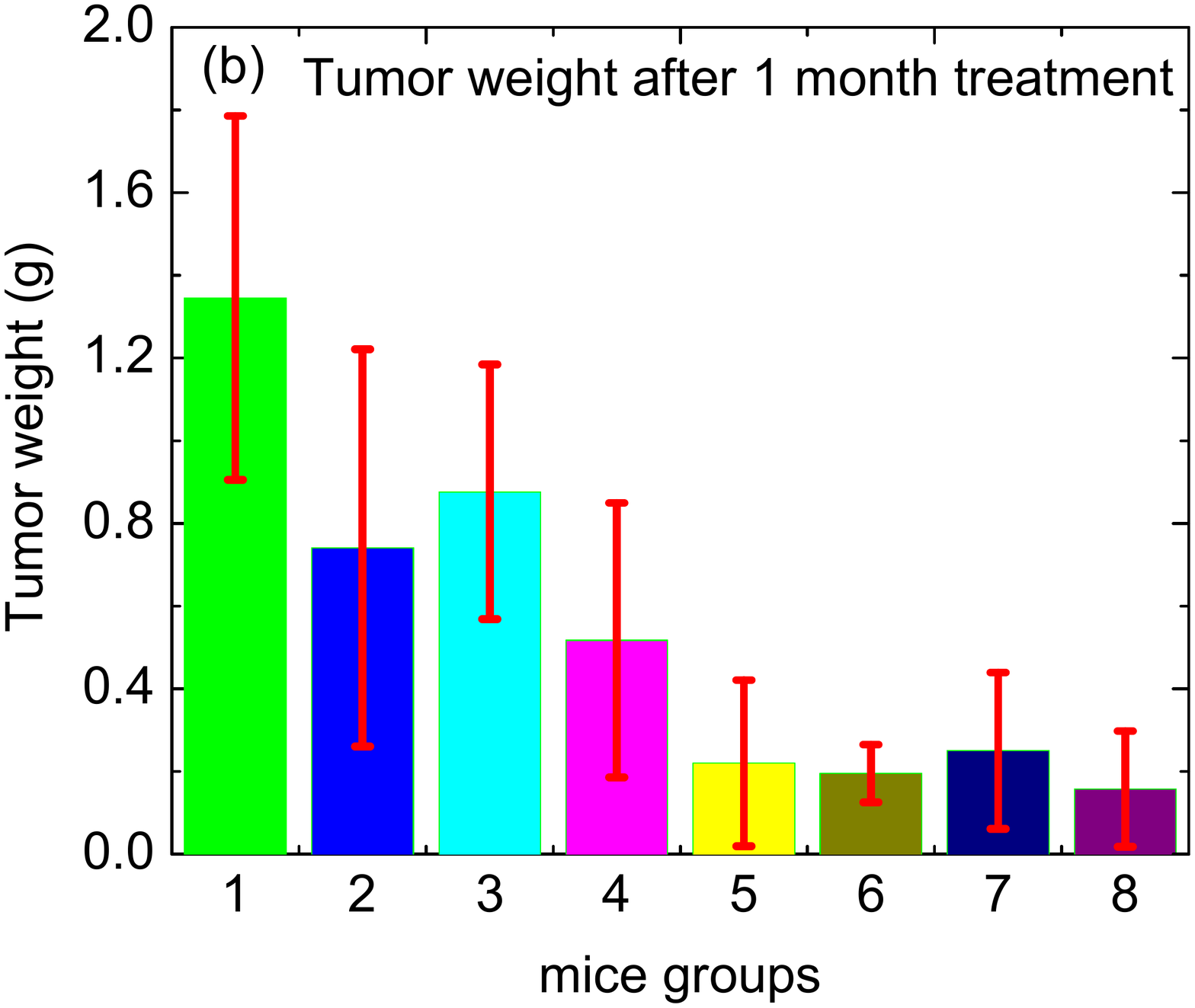}
\caption{\label{fig:8}(Color online) (a) Photograph of tumors after excision from 8 groups. (1) untreated group, (2) laser only, (3) treatment with \ce{GNR$@$SiO_2} OD = 9.2, (4) treatment with \ce{GNR$@$SiO_2}-DOX OD = 9.2. (5) and (6) treatment with \ce{GNR$@$SiO_2} OD = 5.5 and 9.2, respectively, (7) and (8) treatment with \ce{GNR$@$SiO_2}-DOX OD = 5.5 and 9.2, respectively, and under NIR laser irradiation (808 nm, 3.25 \ce{W/cm^2}, 1 minute). (b) The average tumor weights of each group.}
\end{figure}

The chemo- and photothermal therapies in mice due to the use of \ce{GNR$@$SiO_2} and \ce{GNR$@$SiO_2}-DOX are also investigated by the implanted mouse tumor models as described in Section II. C. Figure \ref{fig:8}a shows photographs of tumors taken from 8 animal groups after 30 days treatment. We measured the average tumor weight of each group. The results are shown in Figure \ref{fig:8}b to quantify contribution of chemotherapy and photothermal therapy. Overall, the presence of \ce{GNR$@$SiO_2} and \ce{GNR$@$SiO_2}-DOX without laser exposure destroys tumor cells. Treatment efficiency of using laser illumination on tumors without plasmonic nanostructures is better than just only using \ce{GNRs$@$SiO_2} without laser illumination. However, coating DOX molecules onto the plasmonic composites (group 4) approximately reduces the tumor volume and weight by a factor of 1.5 and 2.7 compared to only \ce{GNR$@$SiO_2} (group 3) and saline injection (group 1), respectively. By turning on the NIR laser light when tumors have \ce{GNR$@$SiO_2}-DOX, we combine effects of the anti-cancer drug and photothermal agents. This combination significantly enhances the efficiency and effectiveness of \ce{GNR$@$SiO_2}-DOX dose. Using more photothermal agents enables to absorb more optical energy to thermally devastate tumor tissue. One can deduce this conclusion from comparing group 7 and 8, which correspond injections with the optical density of 5.5 and 9.2, respectively.
\section{Conclusions}
We have validated anti-cancer effects of the chemo-photothermal therapy based on \ce{GNR$@$SiO_2}-DOX by \emph{in vivo} experiments. We have experimentally prepared \ce{GNR$@$SiO_2}-DOX and examined complex structures via SEM and TEM imagine, and XRD technique. Based on absorption spectra, one can observe variations of plasmonic properties of GNRs of size 10.7 nm $\times$ 39.2 nm when coating a mesoporous silica layer on the gold surface and encapsulating DOX molecules onto this silica layer. Moreover, these optical spectra allow us to determine the amount of loaded DOX molecules. After injecting solutions of the \ce{GNR$@$SiO_2} and \ce{GNR$@$SiO_2}-DOX into mice and scarifying them, photothermal experiments are carried out in both water and livers of mice. These experiments proves that \ce{GNRs$@$SiO_2} are accumulated in livers. The mice liver can be approximately considered as a homogeneous medium and the injected GNRs are randomly dispersed in the liver when injected into mice. When we inject solutions of \ce{GNRs$@$SiO_2} and \ce{GNRs$@$SiO_2}-DOX into tumor-implanted mice and irradiate NIR laser light on their tumors within 1 minute immediately after direct injection, health and weight of mice are observed everyday for one month. Then, their tumors are taken to measure weight. All measurements explicitly indicate that both chemotherapy and photothermal therapy significantly reduce tumor volumes. More reduction of volumes is found when combining these two approaches. 

\section*{Conflicts of interest}
There are no conflicts to declare.
\section*{Acknowledgments}
The animal experiment and all \emph{in vivo} studies were performed according to ethical approval (IRB-A-2001) and guidelines of Institutional Review Board in Animal Research at Dinh Tien Hoang Institute of Medicine, Hanoi, Vietnam. 
The authors are grateful for the financial support for this work from the project VAST.CTVL.02/17-18 and the project 103.03-2016.72 of the Vietnam National Foundation for Science and Technology Development (NAFOSTED).

\section*{Author Information}
Corresponding authors: anh.phanduc@phenikaa-uni.edu.vn, halien@iop.vast.vn

\section*{Supporting Information}
Average body weight of nine groups of mice \emph{in vivo} in experiments as a function of time, TEM image of the mesoporous silica coating on GNRs, and histogram of the rod length and width of GNRs.
%\begin{suppinfo}
%This will usually read something like: ``Experimental procedures and
%characterization data for all new compounds. The class will
%automatically add a sentence pointing to the information on-line:
%\end{suppinfo}

%\bibliography{achemso-demo}

\end{document}